\documentstyle[aps,prl,multicol]{revtex}
\input{epsf}
\begin{document}
\draft
\preprint{\today}
\title {Multifractality beyond the Parabolic Approximation:
    Deviations from the Log-normal Distribution at Criticality in
    Quantum Hall Systems}
\author{Imre Varga, J\'anos Pipek}
\address{Department of Theoretical Physics, Institute of Physics,
    Technical University of Budapest, H--1521 Budapest, Hungary}
\author{Martin Janssen, Krystian Pracz}
\address{Institut f\"ur Theoretische Physik, Universit\"at zu K\"oln,
    Z\"ulpicher Str. 77, D-50937 K\"oln, Germany}
%\date{\today}
\maketitle
\begin{abstract}
    Based on differences of generalized R\'enyi entropies nontrivial
    constraints on the shape of the distribution function of broadly
    distributed observables are derived introducing a new parameter in
    order to quantify the deviation from lognormality. As a test
    example the properties of the two--measure random Cantor set are
    calculated exactly and finally using the results of numerical
    simulations the distribution of the eigenvector components
    calculated in the critical region of the lowest Landau--band is
    analyzed.
\end{abstract}
%\vspace{0.5cm}
%\twocolumn
%\newpage
\begin{multicols}{2}
\narrowtext

    Numerical simulations of disordered critical systems always
    face the problem of dealing with system sizes smaller then the
    relevant length scale therefore highly complex fluctuations
    appear. Consequently expectation values of observables do not
    represent the full picture for the description of the dynamics and
    physics involved, instead the whole distribution function of these
    observables should be studied. This problem has become the central
    issue of theoretical research over the past
    decade~\cite{imry,AKL,PMP,ALW}.

    Multifractal analysis~\cite{muf1} has now become a standard tool
    to handle the aforementioned fluctuations arising near
    criticality. Especially the disorder induced
    localization-delocalization (LD) transition is an important and
    widely studied example of such critical systems~\cite{mj}.

    The general method of multifractal analysis starts with the
    calculation of the measure $Q(\ell)$ of the observable over boxes
    of linear size $\ell$ in a system of length $L$ divided into $M$
    pieces of these boxes. In the thermodynamic limit of $\ell/L\to 0$
    the $q$th moment of this distribution scales with a nontrivial
    exponent $D_q$ that is, in general, strongly $q$--dependent. These
    moments are often called as generalized inverse participation
    numbers (IPN). The quantity $(q-1)D_q$, also known as the mass
    exponent $\tau_q$, is a monotonous, nonlinear function of $q$.

\begin{figure}[tbh]
\epsfxsize=3in
\epsfysize=2.25in
\epsffile{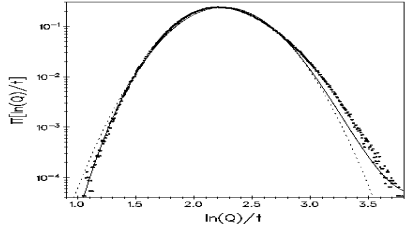}
\vspace{3mm}
\caption[QHS L=200x200]{\label{QHE}
    The joint probability distribution function of the local
    amplitudes $Q=|\psi |^2$ of the eigenstates taken in the critical
    region in a Quantum Hall System. The dashed line represents the
    parabolic approximation, the solid one is the improvement given in
    the text.
}
\end{figure}

    Another way of describing these fluctuations is using the
    $f(\alpha )$ function, the Legendre--transform of $\tau_q$. The
    $f(\alpha )$ is a smooth, single--humped, concave function. It is
    the distribution of the exponents $\alpha_q=d\tau /dq$
    that characterize the $\mu_q(\ell)\sim (\ell/L)^{\alpha_q}$
    scaling of the box--observables where $\mu_q(\ell)$ is the total
    measure of the $q$th power of the observables $Q$ over the boxes
    with linear length $\ell$. The $f(\alpha_q)$ gives the scaling of
    the number of boxes with measure $(\ell/L)^{\alpha_q}$ in the
    interval $(\alpha,\alpha+d\alpha )$
\begin{equation}
    \rho(t,\alpha )=g(t,\alpha )e^{t\,f(\alpha )}
\label{eq:rho}
\end{equation}
    where we have introduced the variable $t=-\ln(\ell/L)$. The above
    distribution describes the multifractal in the sense that
    $\ln\rho(t,\alpha )/t \to f(\alpha )$ if $t\to\infty$.
    The $f(\alpha )$ function obeys certain general rules:
    $f'(\alpha_q)=q$, where prime denotes derivation with respect to
    $\alpha$. For some specific values of $q$
\begin{mathletters}
\label{eq:prop}
\begin{eqnarray}
    f'(\alpha_0)&=&0,\qquad f(\alpha_0)=D_0 \\ \label{eq:propa}
    f'(\alpha_1)&=&1,\qquad f(\alpha_1)=\alpha_1=D_1\label{eq:propb}
\end{eqnarray}
\end{mathletters}
\noindent
    where $D_0$ is the Hausdorff--dimension
    of the support of the observable $Q$. Furthermore the support of
    the $f(\alpha )$ spectrum is a finite interval
    [$\alpha_{-\infty},\alpha_{+\infty}$], and at the two limiting
    values the derivative of the $f(\alpha )$ is infinite. The
    importance of the study of the function $f(\alpha )$ is enhanced
    due to its connection to the properties of the local distribution
    function of the $Q$ variables as~\cite{mj}
\begin{equation}
    \Pi(t,\ln Q)\sim\exp[t(D_0-f(\alpha ))],\qquad \alpha =\ln Q/t
\label{eq:nagypi}
\end{equation}
    The shape of the $f(\alpha )$ curve is close to a
    parabola~\cite{mj,PJ} which means that the distribution $\Pi
    (t,\ln Q)$ is a Gaussian and the $\Pi(t,Q)$ is lognormal.
    Therefore any deviation from the parabolic approximation (PA)
    shows up as a deviation of $\Pi(t,Q)$ from lognormality.
    Numerical experiments show that there is a substantial deviation
    of the distribution function (\ref{eq:nagypi}) from the simple
    Gaussian especially for large values of $\ln Q$, hence it is even
    asymmetrical~\cite{evang-me} (see Fig.~\ref{QHE}).

    The calculation of $f(\alpha )$ and $D_q$ functions in
    numerical experiments suffers from several difficulties. First
    this analysis is suitable only if $\ell\ll L<\xi $ is satisfied,
    where $\xi$ is the localization length and still the box--length
    should be larger that any microscopic length scale of the
    system. Second, higher moments of the distribution, $|q|>3$ are
    usually calculated with low precision and the PA is usually
    satisfactory. The PA is fixed by the values of $\alpha_0$ and
    $D_0$. In spite of the simplicity of the PA it breaks down at
    values of $q$ where the monotonicity of the $\tau_q$ is lost.
% QHS Distribution function
    Apart from the results of numerical simulations, recently it has
    been shown analytically that the moments of the distribution
    function of the eigenstate components in a two dimensional
    disordered system shows multifractality~\cite{falko}, however,
    only the PA has been derived. In another work the multifractal
    properties of a relativistic fermion moving in a two dimensional
    random vector potential field has been derived~\cite{cmw} without
    the restrictions of the PA. Therefore it seems to be of growing
    importance of deriving as much information from numerical
    experiments as possible.

    In this Letter we show that the deviations from the simple
    lognormal distribution can be measured with the application of
    specific R\'enyi--entropies~\cite{renyi}. These R\'enyi entropies
    are proportional to the variable $t$ only for strictly
    self--similar observables in the thermodynamic limit, e.g.
    deterministic multifractals such as the two--measure Cantor set,
    however, they are connected to some specific values of the scaling
    dimensions of the generalized IPNs and hence impose new conditions
    on the shape of the $f(\alpha )$ function.

    Hereby we use two differences of R\'enyi entropies~\cite{renyi}.
    The first one describes the deviation of the
    generalized IPN $\ln P_2$ of the distribution Eq.~(\ref{eq:rho})
    from the Shannon entropy $H$ (a special R\'enyi entropy) and the
    other one is the normalized value of $\ln P_2$
\begin{equation}
    S_2=H - \ln P_2\quad {\rm and} \quad m_2= \ln M - \ln P_2
\label{eq:sq}
\end{equation}
    Generalizations for higher moments are also possible~\cite{pipek},
    however,
    numerical simulations for such cases are presently not too
    reliable. The above two parameters have already been successfully
    applied in a number of systems~\cite{PV1,PV2,VHSP} for the
    shape--analysis of the complex distribution function of
    eigenvector components and energy spacing distributions.

    The calculation of the quantities $S_2$ and $m_2$ involve
    integration over the measure of the distribution in question. In
    our case, we have to perform integration of Eq.~(\ref{eq:rho})
    over the allowed range of the $\alpha$ values
    [$\alpha_{-\infty},\alpha_{+\infty}$]
\begin{mathletters}
\label{eq:int}
\begin{eqnarray}
    P_2^{-1}(\ell)&\sim&\int_{\alpha_{min}}^{\alpha_{max}}
    \ell^{2\alpha}\rho (\ell ,\alpha)d\alpha\\
    \label{eq:int1}
    H(\ell)&\sim& -\ln\ell\int_{\alpha_{min}}^{\alpha_{max}}
           \alpha\ell^{\alpha}\rho (\ell ,\alpha)d\alpha
    \label{eq:int2}
\end{eqnarray}
\end{mathletters}
    Furthermore, being interested only in the small $\ell$, i.e.
    large $t$ limit, we can apply an asymptotic method due originally
    to Laplace~\cite{asymp} and obtain an expansion of $t$ of the
    form~\cite{pipek}
\begin{mathletters}
\label{eq:sm2t}
\begin{eqnarray}
    \label{eq:sm_a}
    S_2(t)&=&(D_1-D_2)t + \sum_{k=0}^{\infty}s_{2,k}t^{-k}\\
    \label{eq:sm_b}
    m_2(t)&=&(D-D_2)t + \sum_{k=0}^{\infty}m_{2,k}t^{-k}
\end{eqnarray}
\end{mathletters}
    From Eqs.~(\ref{eq:sm2t}) it is clear that their ratio tends to a
    constant as $t\to\infty$
\begin{equation}
    \gamma_2 = \lim_{t\to\infty} {{S_2(t)}\over{m_2}(t)} =
                    {{D_1-D_2}\over{D_0-D_2}}.
\label{eq:gam}
\end{equation}
    This is the key result of our Letter. The $D_1$ and $D_2$ values
    in Eqs.~(\ref{eq:sm2t},\ref{eq:gam}) are related to the $f(\alpha
    )$ function, since (\ref{eq:propb}) $\alpha_1=f(\alpha_1)=D_1$ and
    $D_2=2\alpha_2-f(\alpha_2)$. On the other hand one can calculate
    the quantities $S_2/m_2$ directly using the observable $Q$ at any
    length scale. The latter value imposes a further condition on the
    $f(\alpha)$ through the dimensions $D_1$ and $D_2$ using relation
    Eq.~(\ref{eq:gam}): $f(\alpha_2) = 2\alpha_2 - (\alpha_1-\gamma_2
    D_0)/(1-\gamma_2)$. Note that for a lognormal
    distribution $\gamma_2=1/2$ independently of the details of the
    distribution as it is also $\gamma_2=1/2$ for a parabolic
    $f(\alpha )$, hence a deviation from the value 1/2 is a sign of
    non-lognormal distribution and non-parabolic $f(\alpha
    )$~\cite{imi}. Furthermore, the denominator of Eq.~(\ref{eq:gam})
    is the quantity that has been recently connected to the anomalous
    diffusion exponent by Refs.~\cite{PJ} and \cite{chalker}.

    As an illustration first let us consider the two-measure Cantor
    set representing a binary process~\cite{muf1}. In this
    construction one divides the $[0,1]$ interval into two equal parts
    receiving different, $p$ and $1-p$ measures. While iterating this
    procedure for each subhalf of the interval we get a multifractal
    in $D_0=1$. This is a strictly selfsimilar multifractal if we keep
    the $0\leq p\leq 1/2$ value fixed, because for the $n$th stage
    $S_2^{(n)}=n\,S_2^{(1)}$ and $m_2^{(n)}=n\,m_2^{(1)}$
    where $S_2^{(1)}$ and $m_2^{(1)}$ are the values for the initial
    configuration. Hence all the $s_{2,n}$ and the $m_{2,n}$ in
    Eq.~(\ref{eq:sm2t}) are zero, and the $\gamma_2$ quantity is
    independent of $n$ and is a function of $p$ \cite{pipek}.

\begin{figure}[tbh]
\epsfxsize=3in
\epsfysize=2.25in
\epsffile{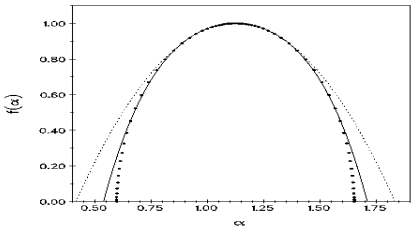}
\vspace{3mm}
\caption[Random Binary Cantor Set]{\label{RBCSFig}
    The distribution of the random binary Cantor set (RBCS) with
    $\delta=1/3$. The dashed line stands for the parabolic
    approximation and the solid one is the fourth order one using
    parameter $\gamma_2$ (see. Eq.~(\protect\ref{eq:gam})). The
    symbols represent the exact relation.
}
\end{figure}

    Next we consider a generalization of this Cantor set by choosing
    at every stage a different value for $p_n$ then we obtain a
    non--selfsimiliar multifractal the random binary Cantor set
    (RBCS). It is straightforward to derive the $f(\alpha )$ function
    once the distribution function of the $p_n$ sequence is further
    specified. Taking a uniform distribution over the
    interval $[1/2-\delta ,1/2+\delta ]$ we have found that
    $\gamma_2(\delta)$ is a smooth function of $0\leq\delta\leq 1/2$.
    Using this simple mathematical construction we will show
    that the PA can be easily improved with the application of
    parameter $\gamma_2$ Eq.~(\ref{eq:gam}). As an example we took
    $\delta=1/3$. The PA in Fig.~\ref{RBCSFig} is denoted as a dotted
    line while the exact relation is represented by the solid symbols.
    We can see that adding a fourth order term $(\alpha-\alpha_0)^4$
    to the PA fulfilling both Eqs.~(\ref{eq:prop}) and (\ref{eq:gam})
    in $f(\alpha)$ gives the solid line that is clearly a much better
    approximation than the simple PA.
% Random Binary Cantor Set
    Finally we will apply our analysis to the widely studied LD
    transition that takes place in a two dimensional disordered system
    subject to a perpendicular strong magnetic field: the quantum Hall
    system~\cite{mj,qhe}. This LD transition is responsible to the
    integer quantum Hall effect. It is described by the critical index
    $\nu$ of the localization length and the $f(\alpha )$ spectrum,
    that is believed to be universal \cite{mj}, however, up to now
    the focus has been set on the PA only and the value of $\alpha_0$
    fixing the position of the maximum of it. The value of $\alpha_0$
    describes the scaling behavior of the typical local electron
    density. We show how the deviation from the lognormal distribution
    can be measured using Eq.~(\ref{eq:gam}) and propose an optimal
    modification of both the $f(\alpha )$ and the distribution
    function $\Pi (t,\ln Q)$. For that purpose we have performed
    calculations of two dimensional systems of linear size of 200
    magnetic lengths. We have obtained 134 eigenfunctions and
    calculated the overall joint probability distribution function of
    the local amplitudes $Q=|\psi |^2$. In Fig.~\ref{QHE} we have
    plotted the logarithm of the histogram of $\ln Q$ as a function of
    $\ln Q/t$. On the same figure the PA is also presented. The
    discrepancy is clear especially for larger values of $Q$. On the
    same figure we have given a possible improvement of the analytical
    $f(\alpha)$ that shows a much better resemblance to the numerical
    data.
% Pade approximation
\begin{figure}[tbh]
\epsfxsize=3in
\epsfysize=2.25in
\epsffile{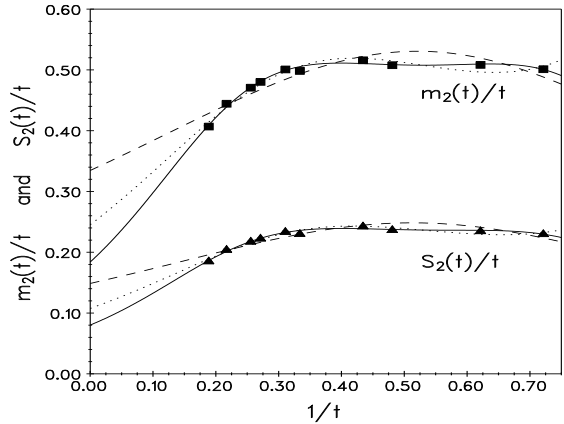}
\vspace{3mm}
\caption[Pad\'e approximation]{\label{Padefig}
    Scaling dependence of parameters $m_2(t)/t$ and $S_2(t)/t$ as a
    function of $1/t$. Solid symbols stand for the values obtained in
    the Quantum Hall System and continuous curves show several [$K/L$]
    order Pad\'e approximants: dashed line [0/2], dotted line [0/3]
    and solid line [0/4].
}
\end{figure}
    The improvement in Fig.~\ref{QHE} was achieved similarly as for
    the case of the RBCS. For that purpose we have calculated the
    $m_2(t)/t$ and $S_2(t)/t$ values and plotted as a function of
    $1/t$. Unfortunatelly the low system size is a severe limitation
    on our problem, since it is desired to obtain the values of
    $m_2(t)/t$ and $S_2(t)/t$ for $t\to\infty$. However, it is not
    possible to go beyond $t_{max}=\ln L$ and current computation
    possibilities impose a limit of $t_{max}\approx 6.0$. In order to
    obtain an acceptable value for the values $m_2(t)/t$ and
    $S_2(t)/t$ as $1/t\to\infty$ we have performed an [$K/L$] order
    Pad\'e approximation~\cite{pade} for both quantities. Besides
    finite size scaling this type of approximation is ready to give an
    estimate of the coefficients in the expansions Eq.(\ref{eq:sm2t}).
    Not having data obtained for different system sizes at hand we can
    show that the Pad\'e approximation indeed gives acceptable
    results, as well. We assumed that the [$K/L$] order approximation
    should have no singularity in the given interval and has to fall
    to zero for $1/t\to\infty$ since on the scale $\ell\approx L$
    the distribution is homogeneous, thus reducing to the
    possibilities to $K<L$ and $K=0$. In Fig.~\ref{Padefig} we have
    plotted the calculated quantities together with several Pad\'e
    approximants. The limit $1/t\to 0$ gives the approximate value
    for the parameter $\gamma_2$. This way we obtained 0.445 for [0/2]
    0.439 for [0/3] and 0.436 for [0/4]. Therefore we set the value of
    $\gamma_2\approx 0.44$. On the other hand, using standard
    multifractal analysis, our data gives $\alpha_0=<\ln Q>/t=2.24$,
    $D_1=1.78$ $D_2=1.61$ that are all well in the generally accepted
    range~\cite{mj,qhe}. The latter values yield $\gamma_2\approx
    0.44$ again.

    Having $\alpha_0$ and $\gamma_2$ at hand we may extend the
    functional form of $f(\alpha)$ beyond the PA in order to get a
    better aggreement with the numerical distribution function (see
    Eq.~(\ref{eq:nagypi})). For the improvement of the functional form
    of the $f(\alpha)$ entering in Eq.~(\ref{eq:nagypi}) there should
    be a number of possibilities. We used the addition of the term
    $(\alpha-\alpha_0)^3$ and $(\alpha-\alpha_0)^5$ to the PA fixing
    their coefficients requiring the constraints Eqs.~(\ref{eq:prop})
    and (\ref{eq:gam}) within an error of 10$^{-4}$. Therefore the
    $\Pi (x)$ function seems to be modified as
    $\Pi (x)=\Pi_{\rm PA}(x)\,\exp [-t(c_3 x^3 + c_5 x^5)]$
    where $x=\ln Q/t-\alpha_0$, $c_3\approx -0.03$ and $c_5\approx
    0.12$.

%\section{CONCLUSIONS}

    In this Letter we have introduced the differences of R\'enyi
    entropies in order to describe the behavior of any multifractal as
    a function of $t=-\ln (\ell/L)$. These parameters enable to look
    beyond the parabolic approximation and obtain further constraints
    on the shape of the $f(\alpha)$ function as well as the
    distribution function of the multifractal quantity. In order to
    get insight in the $t$ dependence we have applied Laplace's method
    and tried the results on exactly known deterministic and random
    multifractals. In the former case the self similarity property is
    also obtained while the latter case was a good test how the
    improvement from PA can be done.

    Finally we have analyzed the joint distribution function of the
    local amplitudes of wave functions calculated in the critical
    regime of the Quantum Hall System. We have applied the Pad\'e
    approximation in order to mimic the thermodynamic limit. The
    results show a possible tail of the distribution function that is
    irregular for smaller amplitudes.

%\acknowledgements

    {\it Acknowledgements} One of the authors (I.V.) is grateful for
    the warm hospitality at the Institut f\"ur Theoretische Physik,
    Universit\"at zu K\"oln where part of this work has been
    completed. Financial support from Orsz\'agos Tudom\'anyos
    Kutat\'asi Alap (OTKA), Grant Nos. T7238/1993 and T014413/1994 is
    gratefully acknowledged.

%          ================= REFERENCES =================
%

\end{multicols}
\end{document}